\documentclass[twocolumn,english,prl,superscriptaddress,aps,showpacs,amsmath,amssymb,floatfix,preprintnumbers]{revtex4}
\pdfoutput=1
\usepackage[pdftex]{graphicx}

\usepackage[T1]{fontenc}
\usepackage[latin9]{inputenc}
\usepackage{babel}

\begin{document}

\preprint{HD-THEP-10-03}

\title{Large-scale inhomogeneities may improve the cosmic concordance of supernovae}

\author{Luca Amendola}
\affiliation{Institut für Theoretische Physik, Universität Heidelberg, Philosophenweg
16, 69120 Heidelberg, Germany}
\affiliation{INAF/Osservatorio Astronomico di Roma, V. Frascati 33, 00040 Monteporzio
Catone, Roma, Italy}

\author{Kimmo Kainulainen}
\affiliation{Department of Physics, University of Jyväskylä, PL 35 (YFL), FIN-40014
Jyväskylä, Finland}
\affiliation{Helsinki Institute of Physics, University of Helsinki, PL 64, FIN-00014
Helsinki, Finland}

\author{Valerio Marra}
\affiliation{Department of Physics, University of Jyväskylä, PL 35 (YFL), FIN-40014
Jyväskylä, Finland}
\affiliation{Helsinki Institute of Physics, University of Helsinki, PL 64, FIN-00014
Helsinki, Finland}

\author{Miguel Quartin}
\affiliation{Institut für Theoretische Physik, Universität Heidelberg, Philosophenweg
16, 69120 Heidelberg, Germany}
\affiliation{Instituto de Física, Universidade Federal do Rio de Janeiro, CEP
21941-972, Rio de Janeiro, RJ, Brazil}

\begin{abstract}
We reanalyze the supernovae data from the Union Compilation including the
weak lensing effects caused by inhomogeneities. We compute the lensing
probability distribution function for each background solution described by the
parameters $\Omega_{M}$, $\Omega_{\Lambda}$ and $w$ in the presence of
inhomogeneities, approximately modeled with a single-mass population of halos.
We then perform a likelihood analysis in the space of FLRW-parameters and
compare our results with the standard approach. We find that the inclusion of
lensing can move the best-fit model significantly towards the cosmic concordance
of the flat $\Lambda$CDM model, improving the
agreement with the constraints coming from the cosmic microwave background and
baryon acoustic oscillations.
\end{abstract}


\maketitle

\paragraph{Introduction.}

In the standard approach supernovae (SNe) observations are analyzed
in the framework of homogenous Friedmann-Lemaître-Robertson-Walker~
(FLRW) models. However, the universe is known to be inhomogenous,
showing a distribution of large galaxy clusters and filamentary structures
surrounding much emptier voids of size $\approx10-100$ Mpc.
A known effect of these structures on any set of standard candles is weak
gravitational lensing~\cite{Bartelmann:1999yn}. Weak lensing can
cause either brightening or dimming of the source depending on whether
the matter column density along the line of sight is larger or smaller
than the FLRW value.

The fundamental quantity describing this statistical magnification
is the lensing probability distribution function (PDF). The lensing
PDF is specific both to the given FLRW model, and to the particular
spectrum of inhomogeneities introduced.
It is not currently possible to extract the lensing PDF from the observational data
and we have to resort to theoretical models.
Two possible alternatives have been followed in the literature.
A first approach (e.g.~Ref.~\cite{NBPDF}) relates a ``universal'' form of the lensing PDF to the variance of the convergence,
which in turn is fixed by the amplitude of the power spectrum, $\sigma_{8}$.
Moreover the coefficients of the proposed PDF are trained on some specific N-body simulations.
A second approach (e.g.~Ref.~\cite{lsum}) is to build a model for the inhomogeneous universe and directly compute the relative lensing PDF, usually through time-consuming ray-tracing techniques.
The flexibility of this method is therefore penalized by the increased computational time.

In this work we use another approach, based on the stochastic modelling
of the inhomogeneities introduced in Ref.~\cite{Kainulainen:2009dw}. This method combines the
flexibility in modelling with a fast performance in obtaining the lensing
PDF.
To compute one lensing PDF, the numerical implementation \texttt{turboGL 0.4} \cite{Kainulainen:2009dw} takes, with an ordinary desktop computer, a time of order of a second.
This speed performance makes it feasible to do an {\it ab initio} likelihood analysis
in the space of FLRW-models endowed with inhomogeneities. In this
letter we will perform such an analysis for the Union SNe Compilation~\cite{Kowalski:2008ez}.

\paragraph{Setup.}

We will treat inhomogeneities as perturbations over the FLRW model
which is parametrized as usual by the present Hubble expansion rate
$H_{0}=100h$ km s$^{-1}$ Mpc$^{-1}$, the present matter density parameter
$\Omega_{M}$ and the present dark energy density parameter $\Omega_{\Lambda}$
and a constant equation of state $w$. We fix the radiation density to $\,\Omega_{R}=4.2\cdot10^{-5}h^{-2}$.
For inhomogeneities we use a ``meatball'' model~\cite{mbTopology} consisting
of randomly placed spherical halos made of ordinary and dark matter.
In principle these halos need not be virialized, and the spherical
symmetry assumption is not very restrictive.
As was explained in Ref.~\cite{Kainulainen:2009dw},
the weak lensing properties of a given universe model can be described by
a set of matter distribution projections ($z$-dependent column densities)
on a small number of independent redshift slices.
For such projections
any local density contrast, such as a long filament seen edge on,
looks roughly like a spherical halo.

Here we use a simple single-mass halo model which is completely parametrized
by the comoving distance between halos $\lambda_{c}$, the halo
proper radius $R_{p}$ and the density profile. We choose the latter
to be the Navarro-Frenk-White profile~\cite{nfw} with a concentration parameter $c\simeq 6.7$ and we assume
that the halos have virialized with a contrast of 200 at a redshift $z_{\rm vir}$,
whereby (for a given $z_{\rm vir}$) the corresponding $R_{p}$ can be taken constant.
The halo mass is related to the comoving density $n_{c}\equiv \lambda_{c}^{-3}\,$
by $\rho_{c}\,\Omega_{M}=M\, n_{c}$. For numerical values we explored the range
$\lambda_{c}=(5.4, \, 9.0, \, 12.6)  \, h^{-1}$Mpc and correspondingly $M = (0.44, \,  2.0, \, 5.6) \, 10^{14}h^{-1}\Omega_{M}\, M_{\odot}$ for $z_{\rm vir} =0.8$, and $z_{\rm vir} = (0,\,0.8,\,1.6)$ for $\lambda_{c}=12.6  \, h^{-1}$Mpc.
The numerical value of $R_{p}$ depends on the background matter density at $z_{\rm vir}$.
For the $\Lambda$CDM model the previous range of $z_{\rm vir}$ corresponds to $R_{p}\simeq (0.9,\,0.7,\,0.5)\, h^{-1}$Mpc.

\paragraph{Lensing.}

The meatball model incorporates quantitatively the crucial feature that photons can travel through voids and miss the localized overdensities. This feature is not present, for example, in ``swiss-cheese'' models where the bubble boundaries are designed to have compensating overdensities. Such models have indeed been shown to have on average little lensing effects~\cite{SCpapers,SClensing}. The key quantity in all our analysis is the lens convergence $\kappa$, which in the weak-lensing approximation is given by
\begin{equation}
    \kappa(z_{s})=\int_{0}^{r_{s}}dr\, G(r,r_{s})\,\delta_{M}(r,t(r))\,.\label{eq:kappa}
\end{equation}
Here $\delta_{M}(r,t)$ is the matter density contrast and
$G(r,r_{s})=\frac{3H_{0}^{2}\Omega_{M}}{2c^{2}}\frac{f_{k}(r)f_{k}(r_{s}-r)}{f_{k}(r_{s})}\frac{1}{a(t(r))}$,
where the functions $a(t)$ and $t(r)$ correspond to the FLRW model, $r_{s}=r(z_{s})$ is the comoving position of the source at redshift $z_{s}$ and the integral is evaluated along the unperturbed
light path.
Also, $f_{k}(r)=\sin(r\sqrt{k})/\sqrt{k},\, r,\,\sinh(r\sqrt{-k})/\sqrt{-k}$
depending on the curvature $k>,=,<0$, respectively.

Neglecting the second-order contribution of the shear, the shift in
the distance modulus caused by lensing is expressed solely in terms
of $\kappa$:
\begin{equation}
    \Delta m(z)=5\log_{10}(1-\kappa(z))\,.\label{eq:dm}
\end{equation}
Equations~(\ref{eq:kappa}-\ref{eq:dm}) show that for a lower-than-FLRW
column density the light is demagnified (e.g., empty beam $\delta=-1$),
while in the opposite case it is magnified.

In Ref.~\cite{Kainulainen:2009dw} a fast and easy way to obtain
the convergence PDF for these meatball models was derived. In short, the
formula for the convergence Eq.~(\ref{eq:kappa}) is replaced by
a discretized probabilistic expression:
\begin{equation}
    \kappa(\{k_{im}\})=\sum_{i=1}^{N_{S}}\sum_{m=1}^{N_{R}}\kappa_{1im}\left({k_{im}}-\Delta N_{im}\right)\,.\label{kappa3}
\end{equation}
Here $\kappa_{1im}$ is the convergence due to one halo, at a comoving
distance $r_{i}$, which the photon path intercepts with an impact
parameter $b_{m}$,
$\kappa_{1im}=G(r_{i},r_{s})\,\int_{b_{m}}^{R(t_{i})}\frac{2xdx}{(x^{2}-b_{m}^{2})^{1/2}}\, \frac{\rho_{i}(x)}{\bar{\rho}_{M}}$,
where $\rho_{i}(x)$ is the local halo density and $\bar{\rho}_{M}$
is the FLRW matter density. In practice one divides the comoving distance
$r_{s}$ to the source and the radius $R$ of the halo into bins of
widths $R\ll\Delta r_{i}\ll r_{s}$ and $\Delta b_{m}\ll R$ and lets
the centers of these bins define the allowed values for $r$ and $b$.
The quantity $k_{im}$ in Eq.~(\ref{kappa3}) is a Poisson random
variable of parameter $\Delta N_{im}=n_{c}\Delta V_{im}$, which gives
the expected number of halos within the phase space volume $\Delta V_{im}=2\pi b_{m}\Delta b_{m}\Delta r_{i}$.
That is, Eq.~\eqref{kappa3} defines a convergence as a function
of a {\em configuration} $\{k_{im}\}$ of halos along an arbitrary
line of sight from the observer to the source. The lensing PDF in
the distance modulus $P_{{\rm wl}}(\Delta m,z_{s})$ is then constructed
from a large sample of random configurations $\{k_{im}\}$ using Eqs.~(\ref{eq:dm})
and (\ref{kappa3}). Note that the expected convergence computed from
Eq.~(\ref{kappa3}) is zero, consistent with photon conservation
in weak lensing, because for a Poisson distributed variable the expected
value coincides with its parameter.

\paragraph{Likelihood function.}

\begin{figure}
\begin{center}
\includegraphics[width=7.8cm]{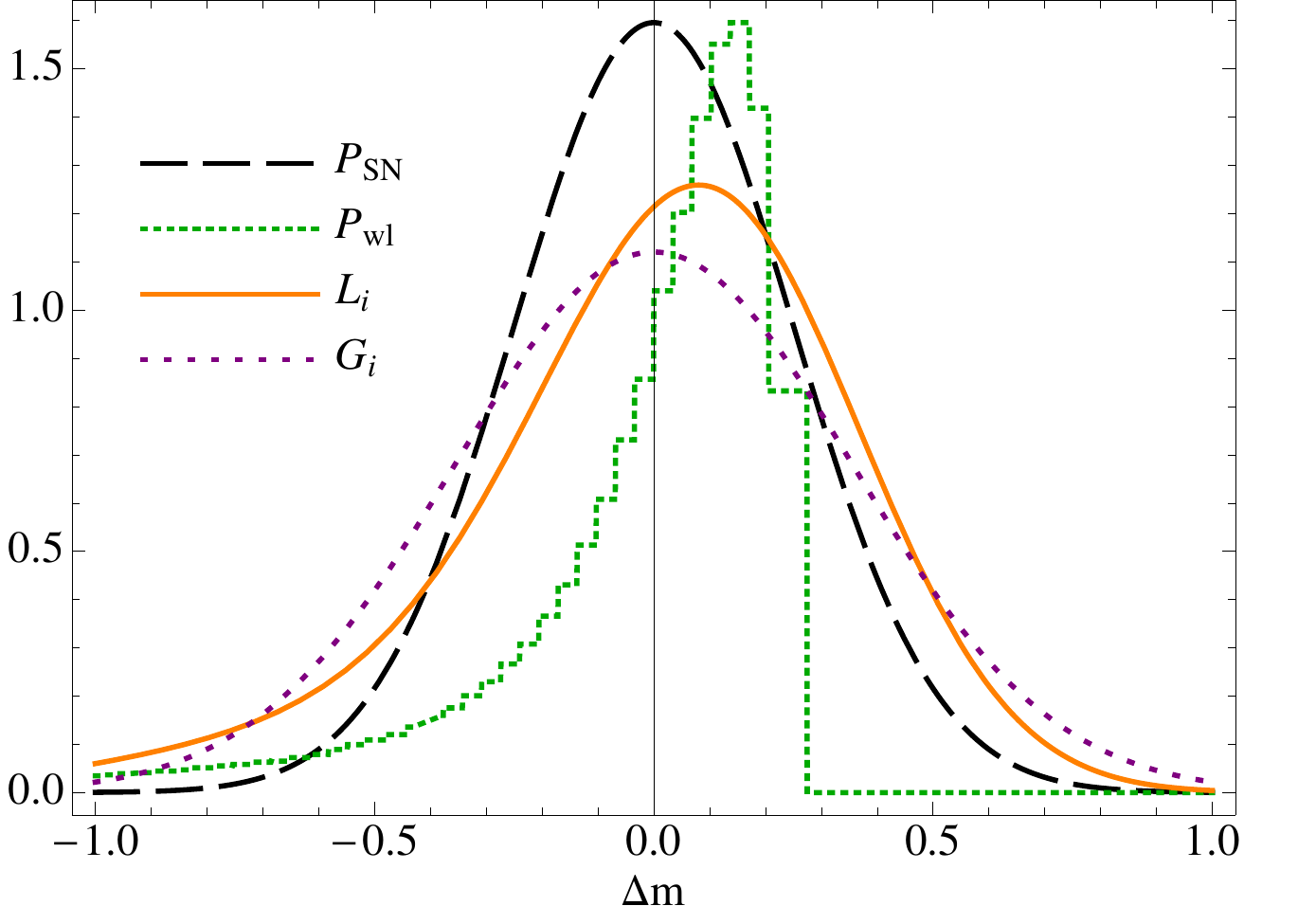}
\caption{PDFs for a SN with $\sigma=0.25$ mag at $z_{s}=1.5$ in the $\Lambda$CDM model endowed with halos specified by $z_{\rm vir} = 0.8$ and $\lambda_{c}=12.6  \, h^{-1}$Mpc. The dotted histogram represents the lensing PDF, the dashed line the SN PDF and the solid line the full likelihood. The dotted curve is described in the text.}
\label{inico}
\end{center}
\end{figure}

After the raw lensing PDF $P_{{\rm wl}}(\Delta m)$ has been computed
for a given set of FLRW-parameters and redshifts, it still has to be
convolved with the intrinsic source brightness distribution $P_{{\rm in}}$:
$P(\Delta m,z_{s})=\int{\rm d}y\, P_{{\rm wl}}(y,z_{s})P_{{\rm in}}(\Delta m-y)$.
We take $P_{{\rm in}}$ to be a gaussian in the distance moduli.
The actual intrinsic distribution should be a universal function if
the SN are similar at all distances. However, following Ref.~\cite{Kowalski:2008ez},
we will combine all observational (gaussian by assumption) uncertainties
in quadrature with the intrinsic distribution, whereby $P_{{\rm in}}$
becomes an effective distribution specific for each SN event $P_{{\rm in}}(x)\rightarrow P_{SN}(x,\sigma_{i})$.
The likelihood function for a single SN-observation is then
\begin{equation}
    L_{i}(\mu)=\int{\rm d}y\, P_{{\rm wl}}(y,z_{i})P_{SN}(\Delta m_{i}-\mu-y,\sigma_{i})\,,\label{LiPDF}
\end{equation}
 where $\Delta m_{i}=m_{o,i}-m_{t,i}$, $m_{o,i}$ is the observed
magnitude and the corresponding FLRW prediction is related to the
luminosity distance $d_{L}$ by $m_{t,i}=5\log_{10}d_{L}(z_{i})/10\,\textrm{pc}$.
The parameter $\mu$ is an unknown offset sum of the SNe absolute
magnitudes, of $k$-corrections and other possible systematics. Note also
that $L_{i}$ inherits the vanishing mean of $P_{{\rm wl}}$
and that its variance is simply given by the sum of the variances of the convolving PDFs.

We define the total likelihood function as the product of all independent
likelihood functions in the data sample, further marginalized over $\mu$:
\begin{equation}
    L(\Omega_{M},\Omega_{\Lambda},w)=\int d\mu\,\Pi_{i}L_{i}(\mu)\,.\label{eq:gbs}
\end{equation}
Since $\mu$ is degenerate with $\log_{10}H_{0}$ we are effectively marginalizing also over the expansion rate of the universe.
A replacement of $P_{{\rm wl}}(y,z)$ by a \textit{cosmology-independent} gaussian with a variance \cite{lsum} $\sigma\equiv0.093z$, would reduce Eq.~(\ref{eq:gbs}) to the form used in the analysis of Ref.~\cite{Kowalski:2008ez}. Typical forms of $P_{{\rm wl}}$, $P_{SN}$ and $L_{i}(\mu=0)$ have been illustrated in Fig.~\ref{inico}. Also shown for later use is  $G_{i}$, which is a gaussian with the same variance of $L_{i}(0)$.

\paragraph{Results.}

\begin{figure}[t]
\begin{center}
\includegraphics[width=7.8cm]{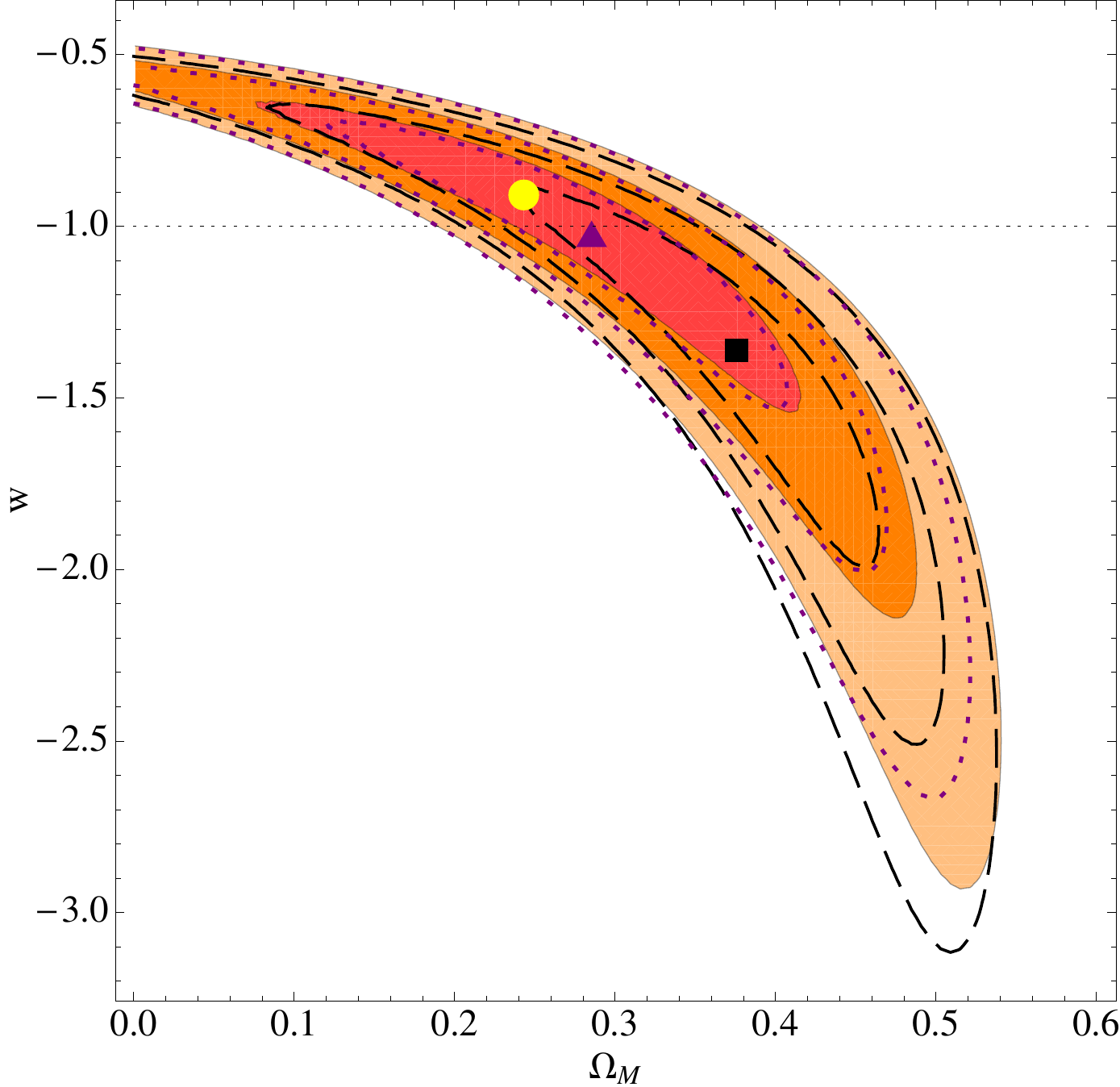}
\caption{1, 2 and 3$\sigma$ confidence level contours on $w$ and $\Omega_{M}$,
for a flat $w$CDM universe with halos specified by $z_{\rm vir} = 0.8$ and $\lambda_{c}=12.6  \, h^{-1}$Mpc.
The results using the full likelihood of Eq.~(\ref{eq:gbs})
are shown as filled contours and the best-fit model with a circle. The
results using the gaussian $G_{i}$ are shown as dotted
lines with a triangle, the ones using the unlensed $P_{SN}$
are shown as dashed lines with a square and correspond to the ones
of Ref.~\cite{Kowalski:2008ez} (without systematics).}
\label{chi2w}
\end{center}
\end{figure}

\begin{figure}[t]
\begin{center}
\includegraphics[width=7.7cm]{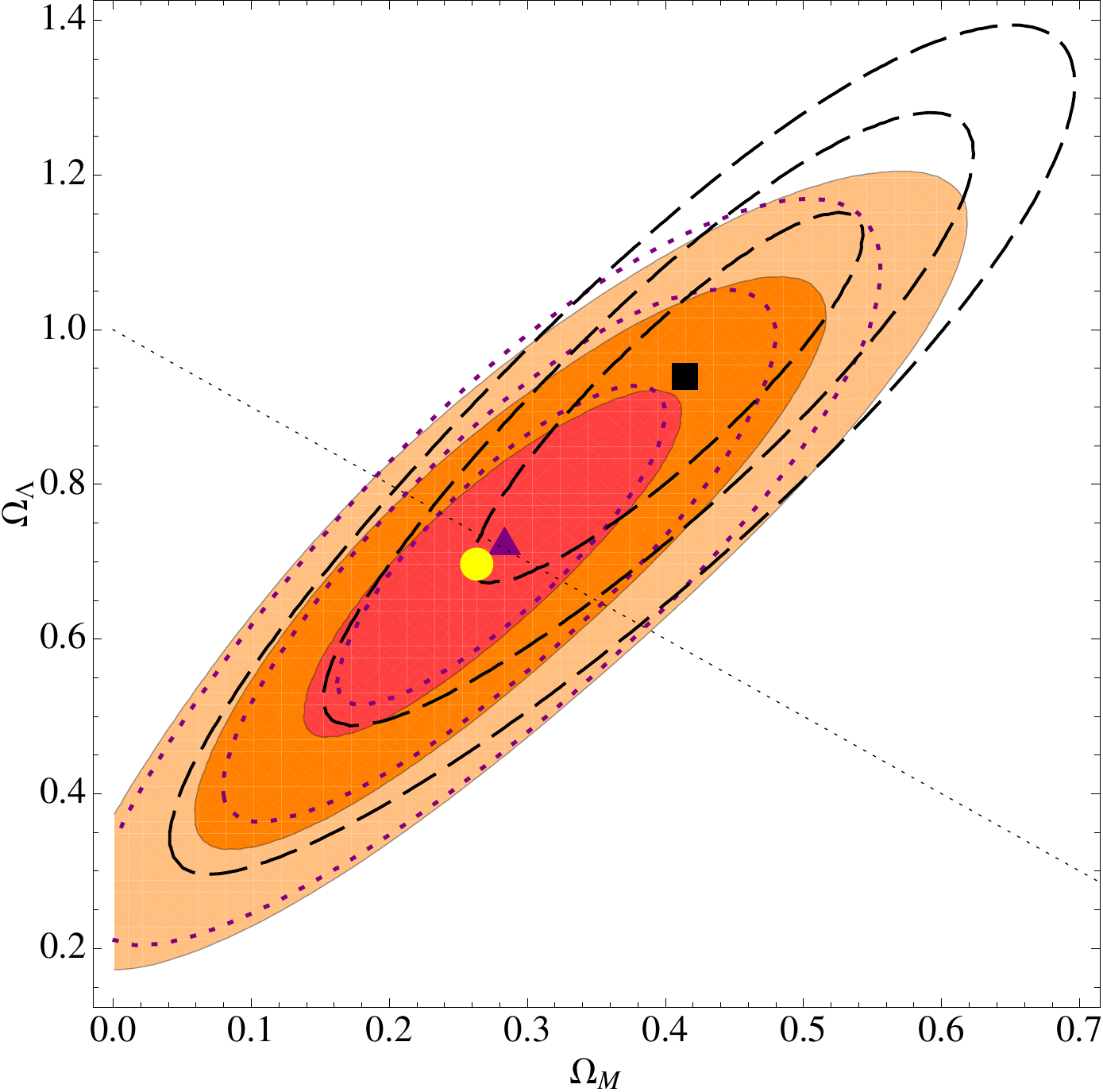}
\caption{1, 2 and 3$\sigma$ confidence level contours on $\Omega_{M}$
and $\Omega_{\Lambda}$ (i.e., allowing for non-zero curvature) for $\Lambda$CDM  ($w=-1$) with halos as in Fig.~\ref{chi2w}.
Note that, as also in the previous plot, the new best-fit points lie on
the 1$\sigma$ confidence level contour relative to $P_{SN}$. Labeling as in Fig.~\ref{chi2w}.}
\label{chi2om}
\end{center}
\end{figure}

We run a global likelihood analysis using the formula (\ref{eq:gbs})
for two different setups: first in the $(\Omega_{M},w)$-space for
flat ($\Omega_{k}=0$) $w$CDM models and second in the ($\Omega_{M},\Omega_{\Lambda}$)-space
for a non-flat $\Lambda$CDM model ($w=-1$) using the Union SNe Compilation of~Ref.~\cite{Kowalski:2008ez}.
We show our results in Figs.~\ref{chi2w} and~\ref{chi2om} as
confidence level contours for $\chi^{2}=-2\log L$. For comparison
we have performed the analysis also using the standard $P_{SN}$ distribution
(as done in Ref.~\cite{Kowalski:2008ez}) and the distribution $G_{i}$.
The idea for using $G_{i}$ is that it takes into account the \textit{cosmology-dependent}
extra dispersion coming from lensing, but neglects the skewness of
the true distribution. So, the contours relative to $G_{i}$ give
an idea of how much of the difference from the standard analysis comes
from the widening of the intrinsic distribution, and how much
from the skewness of the actual PDF. As it is evident from Figs.~\ref{chi2w}--\ref{chi2om}, the 1$\sigma$ contours are basically determined by the cosmology-dependent widening, whereas the skewness starts to be relevant between the 2 and 3$\sigma$ levels.
We point out that our results are essentially unaffected if we add a constat $\sigma_{sys}$ to the $\sigma_{i}$ of Eq.~(\ref{LiPDF}) in order to have the same reduced $\chi^{2}$ using $G_{i}$ and $P_{SN}$.

Our most important result, clearly evident from Figs.~\ref{chi2w}--\ref{chi2om},
is that the inclusion of lensing effects in the likelihood analysis
significantly moves the best-fit model, from $(\Omega_{M}^{V},w^{V})=(0.38,-1.4)$
and $(\Omega_{M}^{V},\Omega_{\Lambda}^{V})=(0.41,0.94)$, towards the cosmic
concordance of the flat $\Lambda$CDM model, therefore improving the
agreement with the constraints coming from cosmic microwave background~(CMB) and baryon acoustic oscillations~(BAO).

To further explore this behavior we studied how the new best-fit model positions $(\Omega_{M}^{*},w^{*})$ and $(\Omega_{M}^{*},\Omega_{\Lambda}^{*})$ depend on the halos mass $M$ and the virialization redshift $z_{\rm vir}$.
For a fixed halo mass, higher values of $z_{\rm vir}$ give denser halos with smaller radius $R_{p}$ and higher lensing corrections to the likelihood.
As explained before, the numerical value of $R_{p}$ depends on the background model and we use the $\Lambda$CDM as a reference model to convert $z_{\rm vir}$ into $R_{p}$. A fit for $\lambda_{c}=12.6  \, h^{-1}$Mpc then gives:
\begin{align}
    (\Omega_{M}^{*},w^{*}) &= (\Omega_{M}^{V}[1-1.5 e^{-2.3 R_{p}}], w^{V}  [1-0.94 e^{-1.7 R_{p}}] ) \nonumber \\
    (\Omega_{M}^{*},\Omega_{\Lambda}^{*}) &= (\Omega_{M}^{V}[1-0.8 e^{-1.2 R_{p}}], \Omega_{\Lambda}^{V} [1-0.54 e^{-1.1 R_{p}}] )
\end{align}
where $R_{p}\geq0.5$ is in units of $h^{-1}\,$Mpc.
If we fix $z_{\rm vir}$, higher values of $M$ (or equivalently $\lambda_{c}$) give a universe made of larger clumps with larger voids giving therefore stronger lensing corrections.
A fit for $z_{\rm vir} = 0.8$ gives:
\begin{align}
    (\Omega_{M}^{*}, \, w^{*}) &= (\Omega_{M}^{V}-0.13 \, M^{0.47}, \, w^{V} +0.45 \, M^{0.33}) \nonumber \\
    (\Omega_{M}^{*}, \, \Omega_{\Lambda}^{*}) &= (\Omega_{M}^{V}-0.15 \, M^{0.29}, \,  \Omega_{\Lambda}^{V} -0.25 \, M^{0.26})
\end{align}
where $M\leq1$ is in units of $5.6 \cdot 10^{14}h^{-1}\Omega_{M}\, M_{\odot}$.

The general trend favoring models with smaller $\Omega_{M}$ follows
from the fact that lensing effects in general make the fit slightly
worse with than without lensing \cite{Kainulainen:2009sx}.
The effect comes both from the cosmology-dependent widening and from the skewness of
the distributions, and it is obviously more pronounced for larger matter
densities. This can be seen directly from Eq.~\eqref{eq:kappa},
where the magnitude of the lensing effects is explicitly seen to be
proportional to $\Omega_{M}$. The overall movement of the best-fit
model then follows the degeneracy of the FLRW models.

\paragraph{Discussion.}

Our halo model was designed to capture the most important effects of the weak gravitational lensing by the nonlinear large-scale structures. In particular the large voids that dominate the late-time universe were imposed in the model by concentrating all matter into halos. Accordingly, we chose the halos to have the mass of a very large cluster, {\em i.e.}~of order $10^{14}h^{-1}M_{\odot}$, which then corresponds to an interhalo distance of order $10\, h^{-1}$Mpc.

Given this result, it is natural to ask if our toy model could also give a reasonable approximation to the observed power spectrum. This is not entirely obvious, because weak lensing and power spectrum probe somewhat different aspects of the inhomogeneities. We found that our single-mass halo model tends to concentrate too much power onto the interhalo distance scale, when compared to the non-linear correction to the $\Lambda$CDM spectrum provided for example by the halo model of Ref.~\cite{Smith:2002dz}.

It will clearly be interesting to improve the modelling of the power spectrum by adopting a more realistic halo distribution function $f(M,z)$, and we plan to pursue this in future work. However, this paper was devoted to explore the extent to which lensing can change the supernovae results and this is best done by adopting the simplest  single-halo model with its few parameters. In any case, the choice of the mass function $f(M,z)$ is not that simple; even after fitting the power spectrum well, an efficient lensing requires modelling the voids and filaments that are described by higher order correlation terms. This can in principle be done in the current approach by introducing additional large scale variations to the background density from which standard halo functions are drawn.

Finally, given a large enough SNe dataset one could in principle measure the lensing PDF.
However, to do this properly one would have to understand the selection effects that could for example cut the high magnification tail of the PDF, sizably biasing the average convergence, variance and skewness.

\paragraph{Conclusions.}

We have presented a reanalysis of the supernovae data from the Union Compilation including the lensing effects caused by inhomogeneities. Unlike in the analysis of Ref.~\cite{Kowalski:2008ez,inhoan}, where the lensing effects are accounted for by adding in quadrature a small $z$-dependent variance to the other statistical and systematic errors, we compute the actual probability distribution functions for each different FLRW-model with a spectrum of inhomogeneities designed to mimic the observed large-scale structures.
In particular, large voids
that dominate the late-time universe are imposed on the model by concentrating all matter into halos of mass of order $10^{14}h^{-1}M_{\odot}$. We found that including inhomogeneities significantly changes the likelihood contours (the likelihood peaks, for instance, move of around $1\sigma$) and clearly improves the concordance of the supernova data with the CMB and the BAO,
which may be used to strengthen the case for the standard $\Lambda$CDM
model.

One should be reminded that our findings could change if other effects caused by large-scale inhomogeneities are introduced, e.g., selection or redshift effects. It also remains to be seen how a more realistic inhomogeneous distribution, providing a better fit to the matter power spectrum, would affect these weak-lensing corrections to the SNe contours.




\begin{thebibliography}{11}
\begin{frenchspacing}

\bibitem{Bartelmann:1999yn}
  M.~Bartelmann and P.~Schneider,
  Phys.\ Rept.\  {\bf 340}, 291 (2001).


\bibitem{NBPDF}
  P.~Valageas,
  Astron.\ Astrophys.\  {\bf 356}, 771 (2000);
  D.~Munshi and B.~Jain,
  Mon.\ Not.\ Roy.\ Astron.\ Soc.\  {\bf 318}, 109 (2000);
  Y.~Wang, D.~E.~Holz and D.~Munshi,
  Astrophys.\ J.\  {\bf 572}, L15 (2002);
  S.~Das and J.~P.~Ostriker,
  Astrophys.\ J.\  {\bf 645}, 1 (2006).



\bibitem{lsum}
  D.~E.~Holz, R.~M.~Wald,
  Phys.\ Rev.\  D {\bf 58}, 063501 (1998);
D.~E.~Holz, E.~V.~Linder,
 Astrophys.\ J.\ \textbf{631}, 678 (2005). 



\bibitem{Kainulainen:2009dw}
  K.~Kainulainen and V.~Marra,
  Phys.\ Rev.\  D {\bf 80}, 123020 (2009).
\texttt{turboGL} is available at: \texttt{http://www.turbogl.org}

\bibitem{Kowalski:2008ez}
M.~Kowalski \textit{et al.},
 Astrophys.\ J.\ \textbf{686}, 749 (2008).

\bibitem{mbTopology} C.~Park, J.~R.~Gott~III, A.~L.~Melott
and I.~D.~Karachentsev, Astrophys.\ J.\ \textbf{387}, 1 (1992);
S.~F.~Shandarin and C.~Yess, Astrophys.\ J.\ \textbf{505}, 12
(1998).


\bibitem{nfw}
 J.~F.~Navarro, C.~S.~Frenk and S.~D.~M.~White, 
 Astrophys.\ J.\ \textbf{462} (1996) 563; 
 J.~F.~Navarro, C.~S.~Frenk and S.~D.~M.~White, 
 Astrophys.\ J.\ \textbf{490} (1997) 493. 


\bibitem{SCpapers} V.~Marra, E.~W.~Kolb, S.~Matarrese and A.~Riotto,
 Phys.\ Rev.\ D \textbf{76}, 123004 (2007); 
V.~Marra, E.~W.~Kolb and S.~Matarrese, 
 Phys.\ Rev.\ D \textbf{77}, 023003 (2008). 


\bibitem{SClensing} N.~Brouzakis, N.~Tetradis and E.~Tzavara,
 JCAP \textbf{0804}, 008 (2008); 
 R.~A.~Vanderveld, E.~E.~Flanagan and I.~Wasserman, 
 Phys.\ Rev.\ D \textbf{78}, 083511 (2008); 
 W.~Valkenburg, 
 JCAP \textbf{0906}, 010 (2009). 



\bibitem{Kainulainen:2009sx}
This effect was observed also in:
K.~Kainulainen and V.~Marra,
Phys.\ Rev.\  D {\bf 80}, 127301 (2009).


\bibitem{Smith:2002dz}
  R.~E.~Smith {\it et al.}  [The Virgo Consortium Collaboration],
  Mon.\ Not.\ Roy.\ Astron.\ Soc.\  {\bf 341}, 1311 (2003).


\bibitem{inhoan} For other studies considering lensing effects see, e.g.:
 D.~E.~Holz, 
 Astrophys.\ J.\ \textbf{506}, L1 (1998); 
 J.~Jonsson, T.~Dahlen, A.~Goobar, C.~Gunnarsson, E.~Mortsell
and K.~Lee, 
 Astrophys.\ J.\ \textbf{639}, 991 (2006); 
 S.~Dodelson and A.~Vallinotto, 
 Phys.\ Rev.\ D \textbf{74}, 063515 (2006); 
 T.~Clifton and P.~G.~Ferreira, 
 JCAP \textbf{0910}, 26 (2009). 


\end{frenchspacing}
\end{thebibliography}
\end{document}